 \documentclass[aps,prl,twocolumn,showpacs,superscriptaddress,floatfix,nofootinbib]{revtex4-1}

\usepackage{amsmath, amssymb, latexsym}
\usepackage{amsfonts}
\usepackage{grffile}
\usepackage{graphicx}
\usepackage{dcolumn}
\usepackage{xcolor}
\DeclareMathOperator{\Tr}{Tr}
\DeclareMathOperator{\Imag}{Im}
\DeclareMathOperator{\Real}{Re}

\begin{document}
	
	\title{Hall conductivity of a Sierpinski carpet}
	
	\author{Askar A. Iliasov}
	\email{A.Iliasov@science.ru.nl}
	\affiliation{Institute for Molecules and Materials, Radboud University, Heyendaalseweg 135, 6525AJ Nijmegen, The Netherlands}
	\affiliation{Key Laboratory of Artificial Micro- and Nano-structures of Ministry of Education and School of Physics and Technology, Wuhan University, Wuhan, China}
	
	\author{Mikhail I. Katsnelson}
	\affiliation{Institute for Molecules and Materials, Radboud University, Heyendaalseweg 135, 6525AJ Nijmegen, The Netherlands}
	
	\author{Shengjun Yuan}
	\email{s.yuan@whu.edu.cn}
	\affiliation{Key Laboratory of Artificial Micro- and Nano-structures of Ministry of Education and School of Physics and Technology, Wuhan University, Wuhan, China}
	\affiliation{Institute for Molecules and Materials, Radboud University, Heyendaalseweg 135, 6525AJ Nijmegen, The Netherlands}
	
	\date{\today}
	
	\begin{abstract}
        We calculate the Hall conductivity of a Sierpinski carpet
        using Kubo-Bastin formula. The quantization of Hall conductivity
        disappears when we increase the depth of the fractal.
        The Hall conductivity is no more proportional to
        the Chern number. Nevertheless, these
        quantities behave in a similar way showing some reminiscence of a
        topological nature of the Hall conductivity. We also study numerically
        the bulk-edge correspondence and find that the edge states become less
        manifested when the depth of Sierpinski carpet is
        increased.
	\end{abstract}

	
	\maketitle

	\section{Introduction}
	
    Fractals were very popular in 1980th and various properties of fractals were
    intensively studied at that time
    \cite{HavAbr1987,Tosatti_1986,Feder_1988}. Most of these works were focused
    on classical fractal systems; at the same time, it turned out that
    quantum properties of fractals are also
    unusual and interesting. For example, fractals have Cantor-like energy
    spectrum, which makes them similar to quasicrystals
    \cite{Domany_1983}.  At that time, studies of quantum
    properties of fractal structures were of purely theoretical
    interest. With recent technological advances, fractals can be
    produced by both
    nanofabrication methods and manipulation of
    individual molecules on metal surfaces \cite{Polini_etal2013,
    Gibertini_etal2009, Shang_etal2015, Kempkes_etal2018}. This enhances the interest
    in a deeper understanding of the field. Recent
    theoretical works concerning quantum effects in fractals consider transport
    properties \cite{Veen2016,Veen2017, SongZhangLi2014}, plasmons
    \cite{Westerhout2018}, Anderson localization \cite{SticAkh2016,
    KosKrzys2017}, topological properties \cite{Brzezetal2018, AgPaiShen2018,
    Pal_etal2019, PaiPrem2019}, and other related topics \cite{Golmankhaneh2019,
    Akal2018, PalSaha2018, NandChak2016, NandPal2015}. The conductivity of fractals is especially interesting, due to possible experimental applications, and definitely deserves more attention.
	
    In a seminal work~\cite{TKNN1982} Thouless et al.
    established that
    the off-diagonal (Hall) conductivity of two-dimensional electron gas in
    magnetic field is proportional to the topological invariant called Chern number and
    is, therefore, closely related to topological
    properties of the system. The derivation rely on
    translational invariance of the system, thus the theory cannot be directly
    applied to quasiperiodic or fractal structures
    lacking the translational invariance.
    For quasiperiodic systems, one can obtain non-trivial topological
    properties of Hall conductivity by looking at the Brillouin zone as a
    non-commutative manifold
    \cite{KrausZilbl2015,Tran_etal2015,Bellissard1992}. The same is true for
    disordered systems \cite{Bellisard1994, Prodan2011}. To our
    knowledge, the relation between Hall conductivity and Chern numbers is still
    an open question for the case of fractals. A
    clarification of this issue could also provide better
    understanding of fracton topological order, another hot subject in contemporary condensed matter physics \cite{Halasz_etal2017,
    Devakul_etal_2019}.

    It is also known for the systems with integer dimensions that quantization of
    Hall conductivity is closely related to the existence of edge
    states, in the form of the so called bulk-edge correspondense \cite{Halperin1982, Hatsugai1993, MongShiv2011}. The terms
    \textit{edge} and \textit{bulk} can be well
    defined for the systems without holes, or, at least, for a system, which has
    an integer dimension and finite number of holes.  A
    fractal has an infinite number of holes, and the distribution of these holes is
    dense. Therefore the difference between edge states and
    bulk states should be carefully checked. This can be
    useful for better understanding of the Hall conductivity of
    fractals. One way to do this is to study various
    approximations to fractals with the holes
    of different scales.

    Chern numbers for a Sierpinski carpet were calculated recently in Ref.
    \onlinecite{Brzezetal2018}. The authors have concluded that the Chern numbers of
    a fractal are still quantized in some energy regions; in
    this sense, fractals still can possess non-trivial topological properties. Ref.
    \onlinecite{Brzezetal2018} also studied Hall conductivity by calculating the
    variance of level spacings and its
    response to an added disorder. Technically, it was done by application
    of random matrix theory. However, the applicability of this approach to
    Sierpinski fractals is questionable, since the latter, strictly
    speaking, are not disordered systems. It was shown that the level-spacing
    distribution for some kind of fractals and quasiperiodic systems can have
    power-law behavior which cannot be described within the framework of the conventional random matrix theory \cite{IlKatSheng2019, MachFuj1986,Hernando2015}. Even if
    one can apply the latter to the spectrum of
    Sierpinski carpet, it is not enough to make definite conclusions on the quantization of
    Hall conductivity. Hence, the variance of
    level spacing distribution is not the most reliable tool to determine
    localization properties and their connection to the Hall
    conductivity.
	
    In this work we examine relations between Hall conductivity and topological
    properties of Sierpinski carpet. In order to do this, we calculate the
    Hall conductivity and quasi-eigenstates for various
    iterations of Sierpinski carpet and investigate how they behave
    when the fractal depth is increased. We also compare our results with the calculated Chern numbers from
    the article \cite{Brzezetal2018}.
	
	\section{The model}	
	
    To study fractal structures, we use the single-orbital tight-binding
    Hamiltonian in the nearest-neighbor approximation, that is, the same model
    as in Refs. \onlinecite{Veen2016,Veen2017}:
	\begin{equation}\label{Eq:TBmodel}
	    H= -t\sum_{\langle ij\rangle} e^{i\phi_{ij}} c^\dag_i c_j \,,
	\end{equation}
    where $c^\dag_i$ creates fermion on a lattice site $i$, and $\langle
    ij\rangle$ denotes the nearest-neighbor sites belonging to the studied
    fractal. The influence of magnetic field is modeled by the standard Peierls
    substitution: $\phi_{ij}=2\pi/\Phi_0 \int^j_i \mathbf{A}\, \mathbf{dr}$,
    where $\mathbf{A}$ is the vector potential and $\Phi_0=hc/e$ is the flux
    quantum. We use Landau gauge $\mathbf{A}=(-By,0,0)$.
	
\begin{figure}[ht!]
     \includegraphics[width=0.9\linewidth]{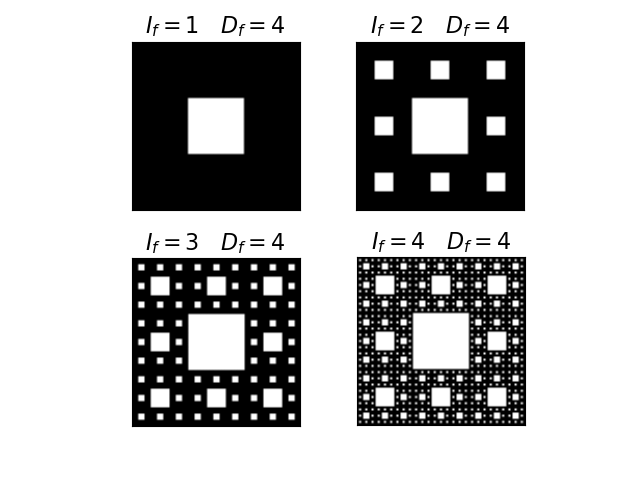}
     \caption{\label{fig:Sierpcarp} The examples of studied fractal
         structures. The size of the large square is $3^{D_f}\times3^{D_f}$ sites
         with $D_f=4$ and different number of iterations $I_f$. With every
         iterations, new holes are deleted. The size of smallest holes is $3^{D_f
         -I_f}\times3^{D_f -I_f}$.}		
	 \end{figure}

    We start with a square lattice of the size $3^{D_f}\times3^{D_f}$. Then, we
    iteratively add holes, which gives us
    the realizations of Sierpinski carpet with varying
    depths. At first, we add the
    central hole of size $3^{D_f-1}\times3^{D_f-1}$, then we
    add holes with size
    $3^{D_f-2}\times3^{D_f-2}$, and so on. We can stop at any number of
    iterations $I_f$ less than $D_f$. The maximum depth of fractal on a given
    lattice is $D_f$. If $I_f = D_f$, the size of a hole is equal to one site. Examples of different iterations are given in Fig. \ref{fig:Sierpcarp}
	
    These structures have two parameters $I_f$ and $D_f$ that describe various
    approximations of Sierpinski carpet. They also describe a
    transition from the usual square lattice with dimension 2, to
    Sierpinski carpet with fractional dimension equal to
    $\ln 8/ \ln 3$.
	
    In order to calculate the density of states and Hall conductivity we use the real-space
    approaches described in Refs. \onlinecite{Hams2000,Shengjun2010,Garcila_etal2015,ShengjunEdo2016}. Using
    these methods, we are able to calculate electronic
    properties for 6 iterations of Sierpinski carpet without any diagonalization.
	
	\section{Results}
	
	\subsection{Density of states}
	
    Let us at first shortly describe the numerical method which we use. We calculate density
    of states (DoS) for different fractal depths and
    iterations using a method based on the time evolution operator
    \cite{Hams2000,Shengjun2010}. We start the evolution of a quantum system
    with a random initial state $|\psi\rangle$, which is normalized so that
    $|\psi|^2=1$. The density of states is calculated via Fourier transform of
    the correlation function by averaging over initial random
    samplings \cite{Hams2000,Shengjun2010}:
	
	\begin{gather*}
	d(\epsilon)=\langle \psi |\delta(\epsilon-H)|\psi\rangle=\\
	=\frac{1}{2\pi}\int^{+\infty}_{-\infty}e^{i\epsilon\tau}\langle \psi|e^{-i\tau H}|\psi \rangle d\tau
	\end{gather*}

    Since density of states is a self-averaged quantity, it
    does not depend on the choice of the state $|\psi\rangle$ for large enough
    systems.

     \begin{figure*}[ht!]
     \mbox{
         \includegraphics[width=0.35\linewidth]{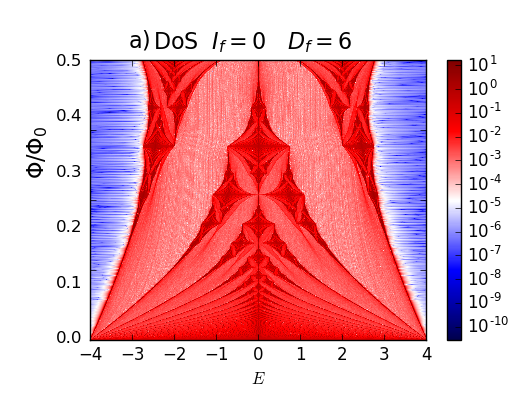}
         \includegraphics[width=0.35\linewidth]{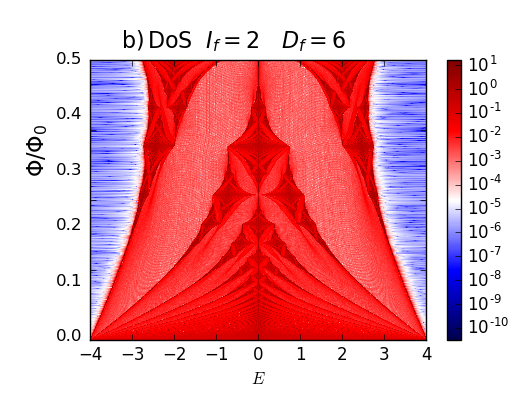}	

     }
     \mbox{
         \includegraphics[width=0.35\linewidth]{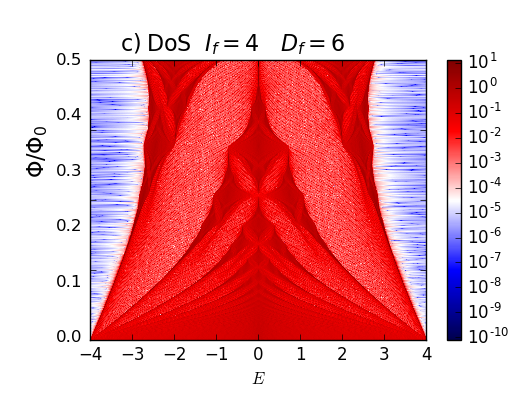}
         \includegraphics[width=0.35\linewidth]{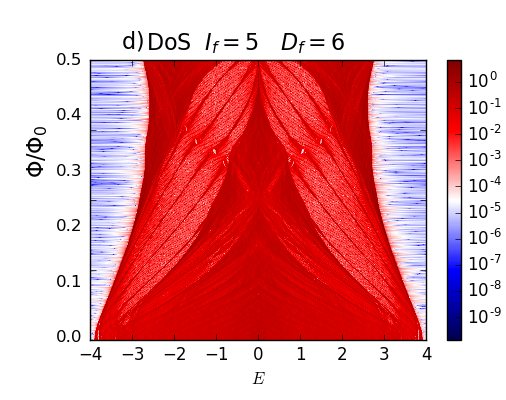}
     }
     \caption{\label{fig:DoS_pictures} Pictures of DoS depending
     on magnetic field -- Hofstadter butterflies (magnetic field
     $B$ corresponds to variations of $\Phi/\Phi_0$ from $0$ to $0.5$) for
     different iterations of Sierpinski carpet in a square of
     size $3^{D_f}\times3^{D_f}$ and $D_f=6$: a)
     shows $I_f=0$ iteration, b) shows
     $I_f=2$ iterations, c) shows $I_f=4$ iterations, d)
     shows $I_f=5$ iterations.
     The differences between a), b) and c) are small. New peaks and gaps appear
     in the picture d).}
	 	
	 \end{figure*}
	
    In Fig. \ref{fig:DoS_pictures}, we show the calculated density of states for
    various magnetic fields corresponding to $\Phi/\Phi_0$ changing from $0$ to
    $0.5$, $\Phi$ is the magnetic flux through the smallest element for a given
    structure. The energy $E$ is measured in values of hopping $t$ of the Hamiltonian
    (\ref{Eq:TBmodel}). These pictures of Hofstadter butterflies
    \cite{Hofstadter1976} are calculated for $D_f=6$ and $I_f=0,2,4,5$.The
    figure demonstrates fractal structure of states and gaps due to additional
    period in hoppings associated with magnetic flux.

    From these pictures one can see that for $I_f=0,2,4$ DoS is basically the
    same for all magnetic fields. The structure of Hofstadter butterfly with
    $I_f=5$ is different from the previous cases. Many more states
    appear and small gaps open for some magnetic fields (new red lines and white dots in Fig.     \ref{fig:DoS_pictures} (d)). Therefore, $I_f=5$ seems to be the minimal depth
    that is needed to catch the peculiarities of quantum states in this particular
    fractal.
		
\begin{figure}[ht!]
	 	\includegraphics[width=0.8\linewidth]{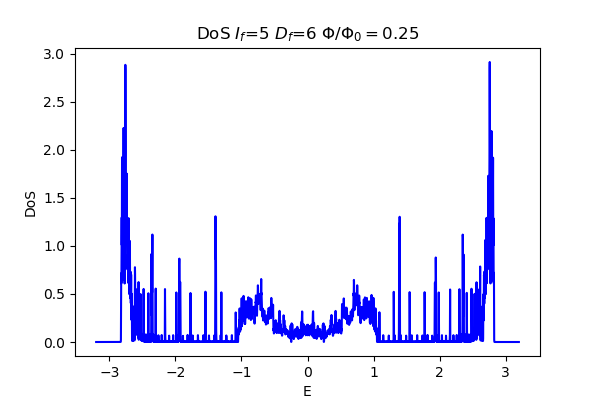}
	 	\includegraphics[width=0.8\linewidth]{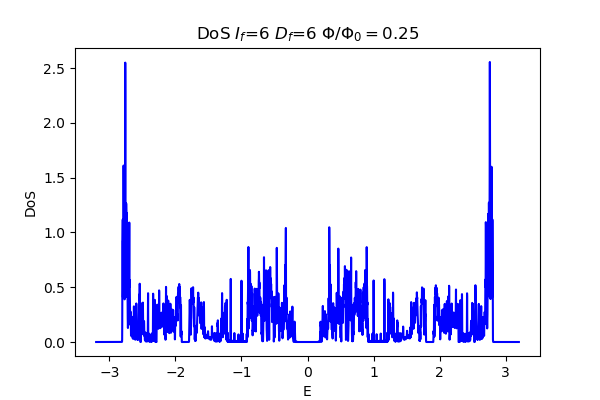}
	 	\caption{\label{fig:DoSN6B1000} The density of states for Sierpinski carpet with $I_f=5$, $D_f=6$  and $D_f=I_f=6$ (maximum number of iterations for the square with size $3^6\times3^6$), $\Phi/\Phi_0=0.25$. A gap appears in the middle of the spectrum for $I_f = 6$, in comparison with $I_f =5$.}		
	 \end{figure}

    In Fig. \ref{fig:DoSN6B1000}, DoS is displayed for fractals with $I_f=5$,
    $D_f=6$ and $I_f=D_f=6$, $\Phi/\Phi_0=0.25$. The change from $I_f =5$ to
    $I_f =6$ is clear: a gap opens in the middle of the
    spectrum, and there are more states between peaks (around energy $E=2$).
   Density of states for $I_f=5$ is flatter in that region.
	
    We also calculated DoS for $I_f=D_f=4$ and $I_f=D_f=5$. The results
    are visually almost indistinguishable from 
    $I_f=D_f=6$. Hence, DoS converges for samples with maximum number of iterations and approaches the thermodynamic limit of a full fractal.
    For cases with $I_f<D_f<6$ we also observe that DoS changes only
    a little while a transition to another regime occurs when going
    from $I_f<D_f -1$ to $I_f=D_f -1$.  Due to this transition, one cannot properly
    approximate the full fractal, if the smallest holes are absent in the sample.

    The occurrence of the transition can be explained by the following
    reasoning.  Before transition point the sample can still be
    treated as a bulk and the effective dimension is an integer. The holes
    in the sample can be seen as some additional
    disorder.  When $I_f$ becomes equal to $5$ the distance between holes
    becomes comparable to the size of a site. Only at this point,
    the effective dimension of a sample becomes non-integer. The difference between cases $I_f = 5$ and $I_f =6$ also can be explained by the difference in
    their non-integer dimensions. Proper approximation
    to Sierpinski carpet is only achieved when
    $I_f = D_f$.

    Let us consider a square with one
    hole $I_f =1$ and then increase the number of sites $D_f$. This
    is an approximation of a two dimensional system. The cases of $D_f=1$ and
    $D_f=2$ correspond to the Sierpinski carpets with $D_f=I_f$ and $D_f=I_f-1$,
    since the average distances between holes are 1 site and 3
    sites. These systems do not behave as two dimensional systems: $D_f=I_f=1$ is just a one dimensional cycle.
    However, in the case of $I_f=1$, for $D_f=3$, the
    sample already has DoS similar to a two dimensional
    sample. Therefore, we assume that samples with $I_f<D_f-1$ approximate
    systems with integer dimensions and transition of physical parameters should happen, when $I_f=D_f-1$.

    We can think about this effect as a crucial property of exact scaling
    symmetry of fractals. Even the smallest
    breakage of scaling symmetry leads to effective integer
    dimension rather than fractional. Every site is an edge
    site in a sample with maximum fractal depth. This condition strongly
    restricts the geometry of paths in a sample. We can assume
    therefore that the scaled geometry plays a decisive role in the properties
    of a fractal and it is closely connected
    to the space of paths in Sierpinski carpet.
	
	\subsection{Hall conductivity}
	
    In order to calculate Hall conductivity, we use the approach from Refs.
    \onlinecite{Garcila_etal2015,ShengjunEdo2016}. The method is based on the so
    called Kubo-Bastin formula:
	
	\begin{gather*}
		\sigma_{\alpha\beta}=\frac{i\hslash e^2}{A}\int^{+\infty}_{-\infty}d\epsilon f(\epsilon)\Tr\langle v_{\alpha}\delta(\epsilon - H)v_{\beta} \frac{dG^{+}(\epsilon)}{d\epsilon}\\-v_{\alpha}\delta(\epsilon - H)v_{\beta} \frac{dG^{-}(\epsilon)}{d\epsilon}\rangle
	\end{gather*}
    where $A$ is the area of the sample,
    $f(\epsilon)=\frac{1}{\exp{(\epsilon-\mu)/T}}$ is the Fermi-Dirac
    distribution, $T$ is the temperature, $\mu$ is the chemical potential,
    $v_{\alpha}$ is $\alpha$ component of velocity operator,
    $G^{\pm}=1/(\epsilon - H \pm i\eta)$ are the Green's functions. In this
    formula we also {average over random samplings as
    we did for the density of states. We can expand Green's functions and delta
    function in Chebyshev polinomials and then for the
    conductivity we obtain:
		
			\begin{equation} \label{eq:Kubocond}
		\sigma_{\alpha\beta}=\frac{4\hslash e^2}{\pi A}\frac{4}{\Delta E^2}\int^{1}_{-1}d\tilde\epsilon \frac{f(\tilde\epsilon)}{(1-\tilde\epsilon^2)^2}\sum_{m, n} \Gamma_{nm}(\tilde\epsilon)\mu^{\alpha\beta}_{m, n}(H)
		\end{equation}
    where $\tilde\epsilon$ is rescaled energy within $[-1, 1]$, $\Delta E$ is
    the energy range of spectrum, $\mu^{\alpha\beta}_{m, n}(H)$ and
    $\Gamma_{nm}(\tilde\epsilon)$ are described by the following formulas:
	
	\begin{gather*}
	\Gamma_{nm}(\tilde\epsilon)=T_m(\tilde\epsilon)(\tilde\epsilon-in\sqrt{1-\tilde\epsilon^2}e^{in \arccos(\tilde\epsilon)})\\
	+T_n(\tilde\epsilon)(\tilde\epsilon+im\sqrt{1-\tilde\epsilon^2}e^{-im \arccos(\tilde\epsilon)})
	\end{gather*}
	
	and
	
	\begin{gather*}
	\mu^{\alpha\beta}_{m, n}(H)=\frac{g_m g_n}{(1+\delta_{n0})(1+\delta_{m})}
	\end{gather*}
    We use Jackson kernel $g_m$ to smooth Gibbs oscillations due to truncation
    of the expansion in Eq. (\ref{eq:Kubocond}) \cite{Garcila_etal2015}.
	
    Our results for $\sigma_{xy}$ are shown in Fig. \ref{fig:Hallcond_pictures}.
    The Hall conductivity was calculated for $\Phi/\Phi_0=0.25$, $D_f=6$ and
    $I_f=0,4,5,6$.

    The Kubo-Bastin formula is derived under very general assumptions and can be
    used to study linear response for any Hamiltonian within single particle approximation. The only restriction of
    the formula is that it neglects the interelectron interactions. For example,
    the Kubo-Bastin formula was successfully applied to systems with
    irregularities, such as disordered systems \cite{Garcila_etal2015}, which
    also lack translational symmetry, and it can be used for fractals as well.
	
    The Hall conductivity behaves similarly to the
    DoS pictures. The differences between $I_f=0$ and $I_f=4$ are quite small,
    although there are fluctuations in the case of $I_f=4$. The structure is
    similar, there are plateaus, which correspond to relatively small values of
    DoS, in the middle of the spectrum and between peaks. These
    plateaus are an indication of relation between Hall conductivity and
    topological invariants: Hall conductivity takes value of $e^2/h$ multiplied by integer (this integer equals to Chern number). One can see a clear transition at
    $I_f=5$ iterations. The plateaus in the middle of the spectrum
    are vanished at $I_f=5$ and fluctuations become much stronger.

    The picture of Hall conductivity for $I_f = D_f = 6$ demonstrates a
    completely different behaviour. We can compare these results to Chern
    numbers calculated in Ref. \onlinecite{Brzezetal2018} (Fig. 3(c) in that
    article) for $I_f = D_f = 4$. In general, Hall conductivity looks similar to
    Chern numbers, however, there are more
    fluctuations and peaks that are absent in the behaviour of Chern numbers. One can also notice that plateaues appear on the scale $1.5 e^2/h$, not $e^2/h$, which would correspond to values of Chern number on these energies.
    The exact plateaus of Chern numbers are located on the energies
    $E=-1.5\ldots-0.9$ and $E=0.9\ldots1.5$, almost quantized region is located
    around $\pm2.5$ with the width of the order of $0.1$. These
    regions are highlighted in Fig. \ref{fig:Hallcond_pictures}(d).
	
    There are two regions which correspond to quantized Chern number, around
    $E=\pm1$. These regions occur after smearing of peaks in
    smaller iteration depths $I_f$. From the previous plateaus
    there remain only small parts around $E=\pm 2.5$ and a part of region around
    $E=\pm 1$, these regions correspond to almost quantized Chern number. It is
    interesting that regions around $E=\pm1$ with quantized Chern numbers are
    not flat plateaus. This was checked for different numbers of random samples
    and fluctuations were stable. The central gap in DoS corresponds to
    conductivity $\sigma_{xy}=0$, as well as the Chern number, which is equal to
    $0$. Therefore, we can conclude that the relation between Chern numbers and
    Hall conductivity does not hold in non-integer dimensions, although there
    are similarities in their behaviour.

    We also added disorder to the sample by deleting random sites. The results
    are shown in Fig. \ref{fig:Hallconddisorder_pictures}, where we deleted
    around $20\%$ of sites in the sample. We see that DoS as well as Hall
    conductivity are stable with respect to the disorder. Moreover, one can see
    even less fluctuations on the pictures in comparison with Fig.
    \ref{fig:Hallcond_pictures}. Thus, we can conclude that fractals are stable
    to disorder, like systems with integer dimensions.

    Particularly this result could be expected, since holes in a fractal sample
    already could be seen as a kind of disorder. Therefore additional disorder
    should not effect physical properties unless this disorder
    is large enough. However, there are subtleties since exact scaling symmetry
    on all scales is important for correct approximation of the fractal.
	
	 \begin{figure*}[ht!]
	 	\mbox{
	 		\includegraphics[width=0.4\linewidth]{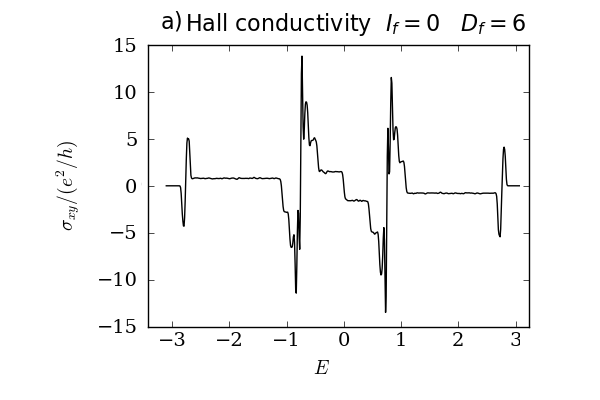}
	 		\includegraphics[width=0.4\linewidth]{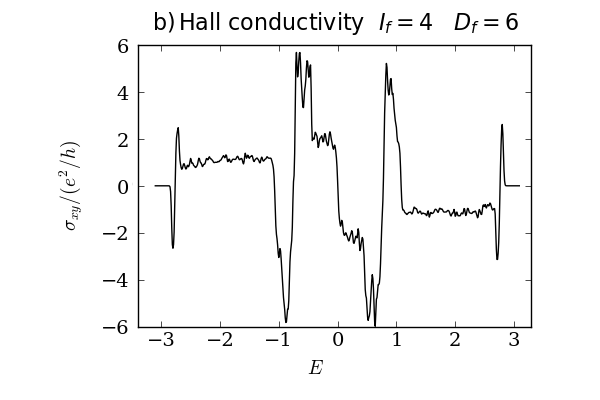}	
	 	}
	 	\mbox{
	 		\includegraphics[width=0.4\linewidth]{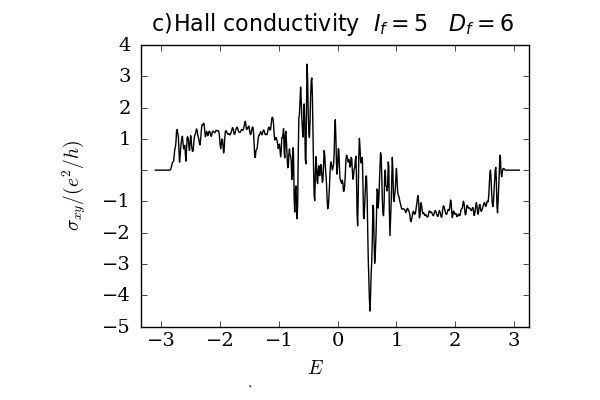}
	 		\includegraphics[width=0.4\linewidth]{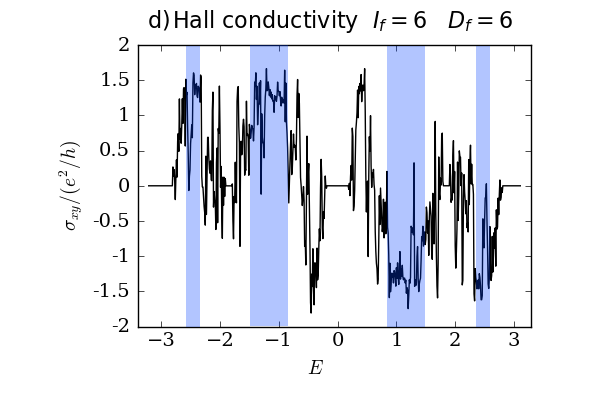}
	 	}
	 	\caption{\label{fig:Hallcond_pictures} The Hall conductivity for different iterations of Sierpinski carpet in a square of the size $3^{D_f}\times3^{D_f}$ and $D_f=6$ ($\Phi/\Phi_0 =0.25$): a) is $I_f=0$ iteration, b) is $I_f=4$ iterations, c) is $I_f=5$ iterations, d) is $I_f=6$ iterations. As in the Fig. \ref{fig:DoS_pictures}, the differences between a) and b) are small, only small fluctuations are added in b). Picture c) demonstrates transition to another phase, picture d) is very different from previous cases, the regions of quantized Chern numbers are shown by blue.}
	 	
	 \end{figure*}

	 \begin{figure*}[ht!]
	 	\mbox{
	 		\includegraphics[width=0.4\linewidth]{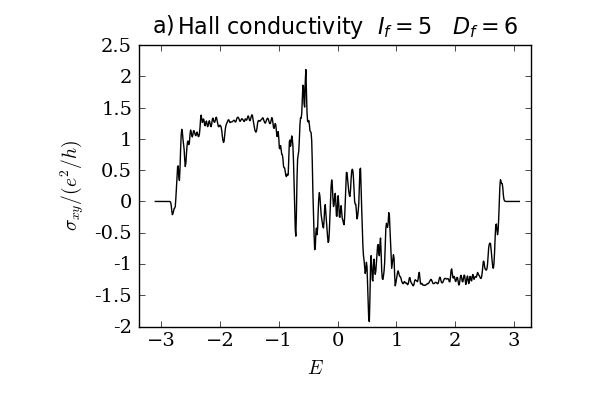}
	 		\includegraphics[width=0.4\linewidth]{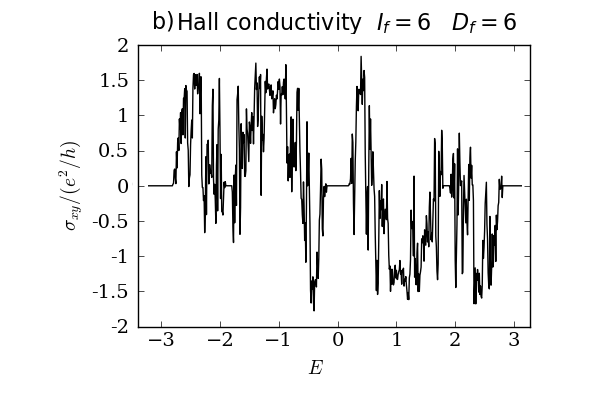}	
	 	}
	 	\caption{\label{fig:Hallconddisorder_pictures} The Hall conductivity for Sierpinski carpet with additional disorder in a square of the size $3^{D_f}\times3^{D_f}$ and $D_f=6$ ($\Phi/\Phi_0 =0.25$): a) is $I_f=5$ iteration, b) is $I_f=6$ iterations. Approximately $20\%$ sites are deleted. There are no visible differences from the Fig. \ref{fig:Hallcond_pictures}}
	 	
	 \end{figure*}

	\subsection{Quasi-eigenstates}

 \begin{figure*}[ht!]
 	\mbox{

 		\includegraphics[width=0.35\linewidth]{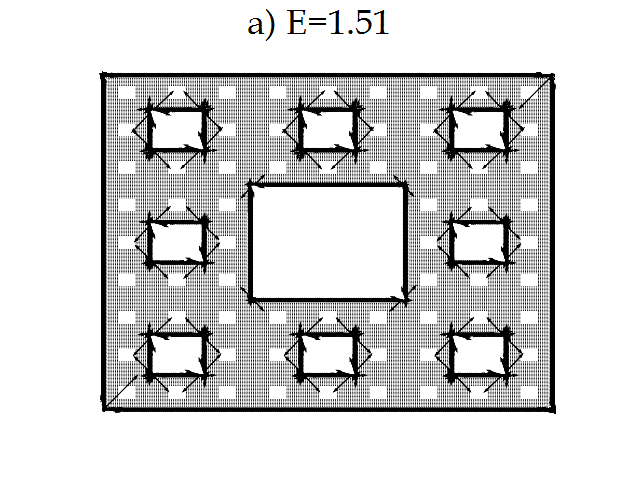}
 		\includegraphics[width=0.35\linewidth]{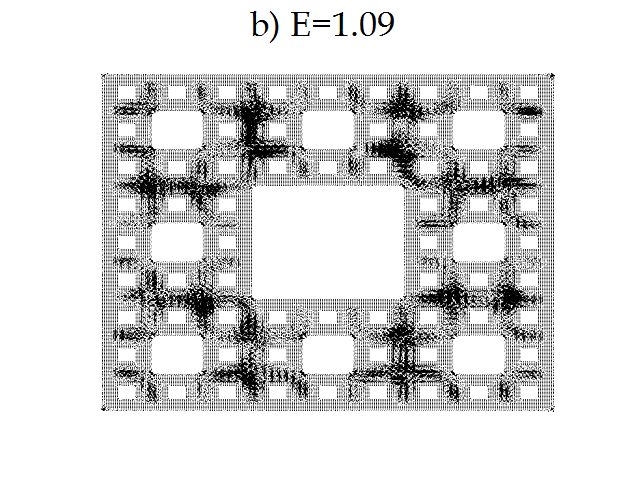}	
 	}
 	\mbox{
 		
 		\includegraphics[width=0.35\linewidth]{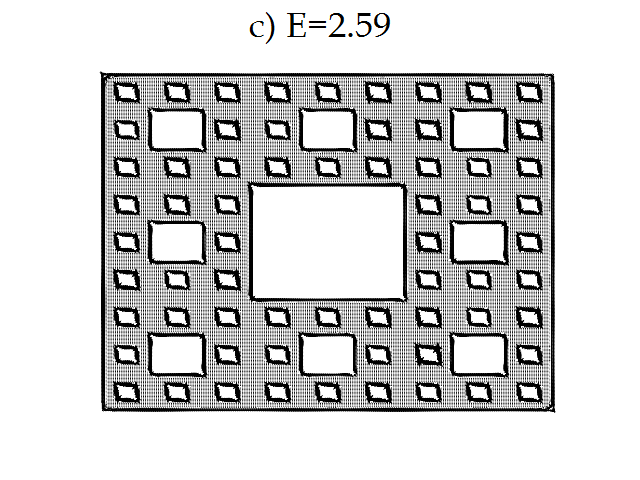}
 		\includegraphics[width=0.35\linewidth]{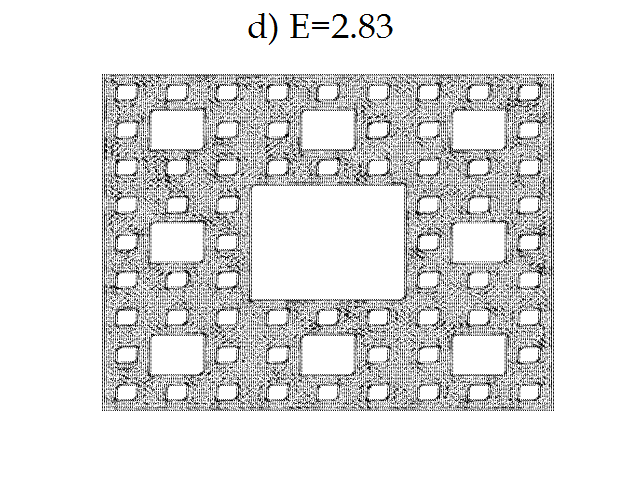}
 		
 	}
 	\caption{\label{fig:states_pictures_3If} Quaisi-eigenstates for Sierpinski carpet with size $3^{D_f}\times 3^{D_f}$  with $D_f=5$ and $I_f=3$ iterations. Examples of edge states are on the left: picture a) $E=1.51$, picture c) $E=2.59$. Examples of bulk states are on the right: picture b) $E=1.09$, picture d) $E=2.83$  }
 	
 \end{figure*}
	
 \begin{figure*}[ht!]
 			\mbox{
 \includegraphics[width=0.35\linewidth]{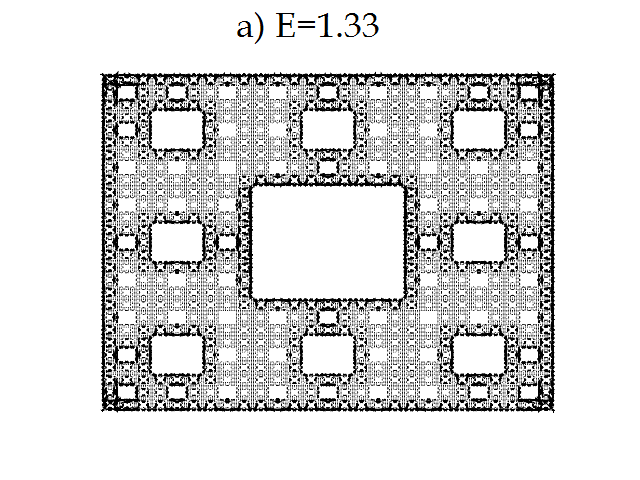}
 		\includegraphics[width=0.35\linewidth]{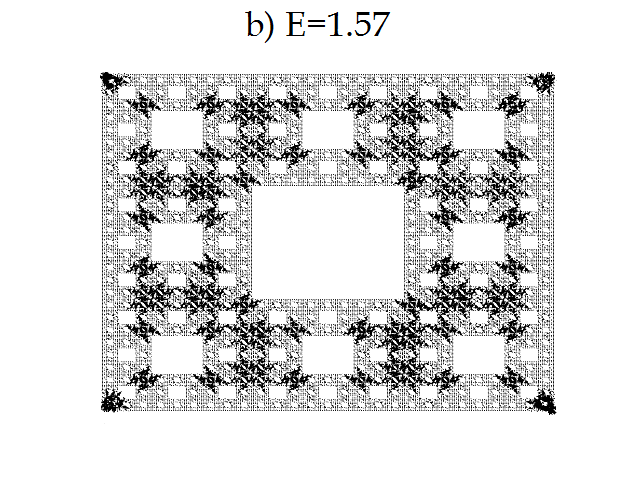}	
 	}
 \mbox{
 	\includegraphics[width=0.35\linewidth]{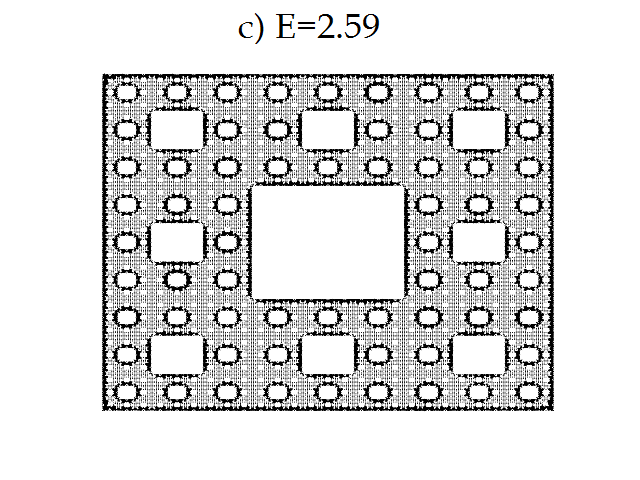}

 \includegraphics[width=0.35\linewidth]{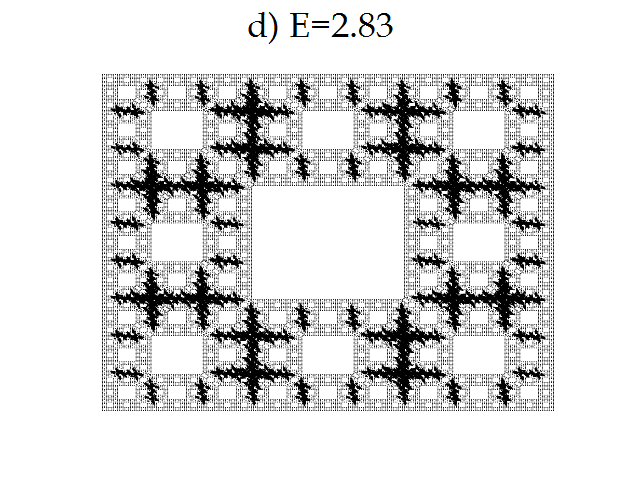}

 }

 	\caption{\label{fig:states_pictures_5If} Quaisi-eigenstates for Sierpinski carpet with size $3^{D_f}\times 3^{D_f}$  with $D_f=5$ and $I_f=3$ iterations. Examples of edge states are on the left: picture a) $E=1.33$, picture c) $E=2.59$. Examples of bulk states are on the right: picture b) $E=1.57$, picture d) $E=2.83$ . 
 	}
 	
 \end{figure*}
	
    It is known that the edge states in quantum Hall systems are closely related
    to their topological properties \cite{Prange1987, Halperin1982,
    Hatsugai1993}. Occurrence of the edge states corresponds to quantized Chern
    numbers (bulk-edge correspondence). Therefore it is
    natural to assume that transitions with increasing
    $I_f$ in DoS and Hall conductivity will be seen in the edge states as well.
    To explore this question we calculated quasi-eigenstates for Sierpinski
    carpet. Quasi-eigenstates and probability current were calculated by the
    same method of averaging \cite{Shengjun2010}. For the probability current,
    we use the formula \cite{Hornberger2002}:
	
	\begin{equation}
	\mathbf{j}=\Real(\psi^{*}\mathbf{v}\psi)=\frac{\hbar}{m_e}\Imag(\psi^{*}\nabla\psi)-\frac{q}{m_e}\mathbf{A}|\psi|^2
	\end{equation}

    First, let us take a look at states
    corresponding to plateaus and peaks for some iterations with enough
     holes, but with regular behaviour of Hall
    conductivity. For more readable pictures we used samples with size $D_f =
    5$. Examples of bulk and edge states for $I_f = 3$ iterations are shown in
    Fig. \ref{fig:states_pictures_3If}. Edge states, which are shown on the left
    side of picture, correspond to energies $E=1.51$ and $E=2.59$. Bulk states,
    which are shown on the right side of picture, correspond to energies
    $E=1.09$ and $E=2.83$. Edge states correspond to plateaus in Hall
    conductivity and bulk states correspond to slopes of peaks. We see that edge
    states can be localized along holes on different scales i.e. only the
    central hole, holes of the second iteration and the central hole, all holes up to the third iteration and so on.

    Examples of bulk and edge states for $I_f = 5$ iterations (i.e. fractal with
    maximum depth) are shown in Fig. \ref{fig:states_pictures_5If}. Edge states,
    which are shown on the left side of the picture, correspond to
    energies $E=1.33$ and $E=2.59$. Bulk states, which are shown on the right
    side of the picture, correspond to energies $E=1.57$ and
    $E=2.83$. Edge states correspond to quantized and almost quantized Chern
    number. There is a reminiscence of this quantization in Hall conductivity.

    We see that the edge states in the full fractal differ from the edge states in Fig.
    \ref{fig:states_pictures_3If}. In the sample with $I_f = 3$ iterations, the
    current is localized along borders in a homogeneous pattern, there are just
    lines of currents along the edges. However, in the full fractal, the current has a more
    complex pattern, for example, it can be localized along small holes, which
    are close to an edge. This deformation of edge current can be a reason of
    absence of plateaus in Hall conductivity even for regions with quantized
    Chern number. We also see that bulk states in a sample with $I_f = 5$
    iterations demonstrate more symmetric behaviour. This is, obviously, the
    manifestation of full scaling symmetry of a fractal.

    We see that for the energies $E=1.33$ and $E=2.59$, which is in
    the region of almost quantized and quantized Chern number, the edge states
    remain to be well-defined. However, some part of plateaus of the
    previous iterations with the edge states become bulk states. We also see that some bulk states in different iterations have similar localization properties.
    Accordingly, we can assume that there are states with different effective
    scales. Some states spread over the rougher structure of a sample, some states sit only around smaller holes.
	
    We have made calculations for various energies and they all
    follow the described pattern. We see that the quasi-eigenstates corresponding to
    the quantized Hall conductivity are localized along the edges
    for iterations smaller than $I_f=4$ (with maximum possible
    number of iterations equal to $D_f=5$). It is also worth
    noticing that all edges, namely, the edges of the sample and the edges of the holes can contribute to quasi-eigenstates. For the most of
    quasi-eigenstates which have been calculated for various energies in the
    case of $I_f=5$ and $D_f=5$, there was no obvious
    domination of edge states.

    We observe that the same transition, which already was observed in DoS and
    Hall conductivity, manifests in quasi-eigenstates, when the number of
    iterations is one less than the possible number of iterations. In the case
    of bulk states, it manifests as a more symmetric picture of
 the current. In the case of edge states, it manifests as
    a more complex localization along the edges, so that
    some edge states become localized along smaller holes, which are close to an
    edge.
	
	\section{Summary}
	
    We see that with increasing the depth of a fractal the quantization of Hall
    conductivity disappears. However, there is some reminiscence of topological
    nature of quantization, namely some plateaus remain. We also see that for
    Sierpinski carpet the relation between topological invariants such as Chern
    numbers and Hall conductivity does not work, contrary to the case of integer
    dimensions. The calculated conductivity is not proportional to the Chern number,
    but follows a similar pattern. One can speculate about
    possible reasons of it.

    At first, the formula, which calculates Chern number through projectors, was
    proven to be properly defined only for systems with translational invariance
    \cite{Kitaev2006}. Even if the formula works for fractals, it may be that one
    needs to calculate Hall conductivity and Chern numbers in the
    thermodynamic limit. Another reason can be related to difficulties with defining edge and bulk states.
    Some of the edge states are localized along inner holes.
    These states become closer to bulk states at maximal
    depth, so we cannot saywhether it is the localization along the edges
    or localization on small inhomogeneities. Edge states along
    large holes also become localized not only along edges, but also along small holes around an edge. These possible effects
    require future investigation.
	
    We considered different iterations $I_f$ of Sierpinski carpet on a fixed
    rectangular sample with size $3^{D_f}\times 3^{D_f}$. We observed that there
    is a transition between two different regimes, which occurs when $I_f=D_f
    -1$. This transition can be seen in density of states, Hall conductivity and
    quasi-eigenstates. In the case of quaisi-eigenstates, the edge states mostly
    become bulk states when $I_f$ is increased. This result can be explained
    by the effective dimension. If the number of
    holes is finite, then the effective dimension of a sample is
    integer rather than fractional. The transition occurs when
    holes in a sample are dense enough and the effective dimension
    of the sample becomes non-integer.
	
    Edge states can be localized along the borders of all
    holes of various scales, not only the edges of the sample.
    There is no big difference in amplitude of currents for different holes. Edge states can still be seen for a full fractal, however, they change their localization
    behavior. Additional holes along the edges increase effective localization width. Therefore, one can speculate that if a state is localized around small
    holes and these holes are dense enough, there could be a transition from
    edge state to a bulk state.

    When this work was finished the preprint \cite{Fritz2019}
    appeared which treated a similar problem but in a technically different way (it was based on Landauer formula rather than Kubo-Bastin formula and
    did not include an analysis of edge states). Qualitatively,
    parts of our conclusions are similar to these obtained in that paper.
	
	\section{Acknowledgements}
We are thankful to Tom Westerhout for the helpful discussions.
This work was supported by the National Science Foundation
of China under Grant No. 11774269 and by the Dutch
Science Foundation NWO/FOM under Grant No. 16PR1024
(S.Y.), and by the by the JTC-FLAGERA Project GRANSPORT (M.I.K.). Support by
the Netherlands National Computing Facilities foundation
(NCF), with funding from the Netherlands Organisation for
Scientific Research (NWO), is gratefully acknowledged.
		
\bibliographystyle{apsrev}

\end{document}